\documentclass[aps,pra,twocolumn,reprint,floatfix]{revtex4-2}
\usepackage[utf8]{inputenc}
\usepackage[T1]{fontenc}
\usepackage{amsmath,amssymb}
\usepackage{graphicx}
\usepackage{booktabs}
\usepackage{hyperref}
\usepackage{xcolor}
\usepackage{float}

\begin{document}

\title{Fermi--Born--Infeld electrodynamics: a nonlinear theory with physical gauge}
\author{Renato Vieira dos Santos}
\email{renato.santos@ufla.br}
\affiliation{Instituto de Ci\^{e}ncia, Tecnologia e Inova\c{c}\~{a}o -- ICTIN, Universidade Federal de Lavras -- UFLA, Campus Para\'{i}so, MG 37950-000, Brazil}

\date{\today}

\begin{abstract}
We construct a nonlinear extension of Fermi's electrodynamics by incorporating a Born--Infeld structure that depends directly on the four-potential $A_\mu$ rather than on the field strength $F_{\mu\nu}$. The resulting theory, which we call Fermi--Born--Infeld (FBI) electrodynamics, eliminates the $U(1)$ gauge redundancy by elevating the Lorenz gauge to a dynamical condition. The Lagrangian is built from the determinant of a metric-like tensor $g_{\mu\nu} = \eta_{\mu\nu} + 2\kappa\, \partial_{(\mu} A_{\nu)}$, ensuring that the canonical energy--momentum tensor and the spin density remain unique and free of gauge ambiguities. We derive the field equations, which reduce to $\partial_\nu(\sqrt{-g}\, g^{\mu\nu}) = 0$, and show that the Lorenz condition $\partial_\mu A^\mu = 0$ emerges dynamically from retarded boundary conditions and the requirement of a positive-energy spectrum. The nonlinearities modify the propagation of longitudinal modes; we argue, via a Vainshtein-like mechanism, that the nonlinear self-interactions may stabilize the longitudinal mode, opening the possibility of a stable massive scalar photon under extreme field conditions. We also compute the spin density from the Noether current and discuss its properties. The FBI theory preserves the physical gauge of Fermi's original formulation while incorporating the regularization features of Born--Infeld electrodynamics, making it a candidate for describing electromagnetic phenomena in strong-field regimes.\\

\end{abstract}

\maketitle

\section{Introduction}
\label{sec:intro}

The quest for a consistent, unique definition of the electromagnetic spin density has motivated a reconsideration of gauge invariance in classical electrodynamics. In the standard Maxwell formulation, the four-potential $A_\mu$ is defined only up to a gauge transformation, leading to ambiguities in the separation of orbital and spin angular momentum \cite{Beth1936,Belinfante1940,Barnett2010,Leader2014}. Fermi proposed an alternative Lagrangian in 1930 that breaks gauge invariance explicitly by adding a term $(\partial_\mu A^\mu)^2$ \cite{Fermi1930,Fermi1932}. In this framework, the Lorenz condition arises as a dynamical consequence of the field equations and retarded boundary conditions, not as an external choice. The resulting theory, which we refer to as Fermi's physical-gauge electrodynamics, provides a unique vector potential and a well-defined local spin density \cite{VanOosten1999}.

The question of a unique local spin density is more than an academic subtlety.
In gauge‑invariant electrodynamics the separation of the total angular momentum of light into orbital and spin contributions is ambiguous, because different gauge choices shift terms between the orbital and spin parts of the Noether current~\cite{Belinfante1940,Leader2014,Barnett2016,Li2017}.
This ambiguity has measurable consequences: the local spin density governs spin‑momentum locking in evanescent waves, optical torques on birefringent particles, and the spin Hall effect of light~\cite{Bliokh2015}.
A formulation that yields an unambiguous, gauge‑free local spin density is therefore of direct experimental relevance.
Fermi's physical‑gauge electrodynamics provides exactly such a formulation, and one of the principal goals of the present work is to extend this feature into the nonlinear regime.

A separate line of development is the nonlinear electrodynamics of Born and Infeld \cite{Born1934}, which was introduced to regularize the self-energy of point charges. The Born--Infeld Lagrangian depends only on the field strength $F_{\mu\nu}$ and is gauge-invariant by construction. Its characteristic square-root form imposes an upper bound on the electric field strength, analogous to a maximum velocity in special relativity. This theory has attracted renewed interest in the context of string theory, where it appears as the effective action on D-branes \cite{Fradkin1985,Leigh1989}, and in high-intensity laser physics \cite{DiPiazza2012}.  For a recent comprehensive review of the nonlinear electrodynamics landscape, see \cite{Sorokin2022}.

In this paper, we combine these two ideas into a single framework: a nonlinear electrodynamics that depends directly on the potential $A_\mu$ rather than on $F_{\mu\nu}$, and that reduces to Fermi's theory in the weak-field limit. We call this theory \emph{Fermi--Born--Infeld} (FBI) electrodynamics. The construction employs a metric-like tensor $g_{\mu\nu} = \eta_{\mu\nu} + 2\kappa\, \partial_{(\mu} A_{\nu)}$, where $\kappa$ is a constant with dimensions of inverse field. The Lagrangian is taken to be proportional to $\sqrt{-\det(g_{\mu\nu})}$, following the Born--Infeld philosophy. Because $g_{\mu\nu}$ is not invariant under gauge transformations of $A_\mu$, the theory has no gauge redundancy; the potential is a physical field determined uniquely by the equations of motion and boundary conditions.

The structure of the paper is as follows. In Sec.~\ref{sec:lagrangian} we construct the Lagrangian and show its expansion to second order, recovering Fermi's theory. Section~\ref{sec:eom} derives the field equations and demonstrates the dynamical emergence of the Lorenz condition. In Sec.~\ref{sec:longitudinal} we analyze the longitudinal modes and discuss their stability under nonlinear effects. Section~\ref{sec:spin} computes the canonical energy, momentum, and spin densities from Noether's theorem. We conclude in Sec.~\ref{sec:discussion} with a summary and outlook. Throughout, we use the SI system of units and the Minkowski metric $\eta_{\mu\nu} = \mathrm{diag}(+,-,-,-)$. Greek indices run from $0$ to $3$, and Latin indices from $1$ to $3$. The Einstein summation convention is employed.

\section{Construction of the Lagrangian}
\label{sec:lagrangian}

\subsection{Motivation and definition}

The standard Born--Infeld Lagrangian for the electromagnetic field is built from the determinant of $\eta_{\mu\nu} + b^{-1} F_{\mu\nu}$, where $b$ is a constant with dimensions of field strength \cite{Born1934}. The result is a gauge-invariant, nonlinear function of $F_{\mu\nu}$ that reduces to the Maxwell Lagrangian at weak fields. To incorporate Fermi's idea of a physical gauge, we replace the antisymmetric $F_{\mu\nu}$ by a symmetric tensor constructed from the derivative of $A_\mu$.

Let $\kappa$ be a constant with dimensions $[A]^{-1}$ (inverse of potential). We define the symmetric second-rank tensor
\begin{equation}
g_{\mu\nu} = \eta_{\mu\nu} + 2\kappa\, \partial_{(\mu} A_{\nu)},
\label{eq:gmunu}
\end{equation}
where $\partial_{(\mu} A_{\nu)} = \frac12 (\partial_\mu A_\nu + \partial_\nu A_\mu)$ is the symmetric part of the gradient. The tensor $g_{\mu\nu}$ has dimensions of $1$ (adimensional) because $\kappa A$ is adimensional. The Fermi--Born--Infeld Lagrangian is then
\begin{equation}
\mathcal{L}_{\rm FBI} = -\frac{1}{\kappa^2 \mu_0} \left[ \sqrt{-\det(g_{\mu\nu})} - 1 \right].
\label{eq:LFBI}
\end{equation}
The factor $\mu_0$ (vacuum permeability) ensures that the Lagrangian has the correct dimensions of energy density. The constant $\kappa^2\mu_0$ has dimensions $[A]^2 \cdot [\mathcal{L}]$, so the overall expression has dimensions $J/m^3$. In natural units where $\mu_0$ is absorbed, one may set $\mu_0=1$; we keep it for SI compatibility.

The choice of the symmetric combination $\partial_{(\mu} A_{\nu)}$ is crucial. Had we used the antisymmetric $F_{\mu\nu}$, the theory would remain gauge-invariant and we would simply recover the standard Born--Infeld electrodynamics in a different representation. The symmetric tensor, by contrast, breaks $U(1)$ gauge symmetry because under $A_\mu \to A_\mu + \partial_\mu \Lambda$, the combination $\partial_{(\mu} A_{\nu)}$ transforms as $\partial_{(\mu} A_{\nu)} \to \partial_{(\mu} A_{\nu)} + \partial_\mu \partial_\nu \Lambda$, which is not a gradient of a vector field and therefore not a pure diffeomorphism of $g_{\mu\nu}$.

\subsection{Weak-field expansion}

For fields satisfying $|\kappa\, \partial A| \ll 1$, we can expand the determinant. Write $g_{\mu\nu} = \eta_{\mu\nu} + h_{\mu\nu}$ with $h_{\mu\nu} = 2\kappa\, \partial_{(\mu} A_{\nu)}$. Then, up to second order in $h$,
\begin{equation}
\begin{split}
&\sqrt{-\det(\eta + h)} = 1 + \frac12 \operatorname{tr}(\eta^{-1}h) \\&+ \frac18 [\operatorname{tr}(\eta^{-1}h)]^2 - \frac14 \operatorname{tr}[(\eta^{-1}h)^2] + \mathcal{O}(h^3).
\label{eq:detexp}
\end{split}
\end{equation}
Using $\eta^{\mu\nu}$ to raise indices, the trace is $\operatorname{tr}(\eta^{-1}h) = \eta^{\mu\nu} h_{\nu\mu} = 2\kappa\, \partial_\mu A^\mu$. The quadratic terms are
\begin{equation}
\operatorname{tr}[(\eta^{-1}h)^2] = h^\mu{}_\nu h^\nu{}_\mu.
\label{eq:trace2}
\end{equation}
With $h^\mu{}_\nu = 2\kappa\, \partial^{(\mu} A_{\nu)}$, we compute
\begin{equation}
h^\mu{}_\nu h^\nu{}_\mu = 4\kappa^2 \, \partial^{(\mu} A^{\nu)} \partial_{(\mu} A_{\nu)} = 4\kappa^2 \, S_{\mu\nu} S^{\mu\nu},
\label{eq:h2}
\end{equation}
where $S_{\mu\nu} = \partial_{(\mu} A_{\nu)}$. Expanding $S_{\mu\nu}S^{\mu\nu}$,
\begin{align}
S_{\mu\nu}S^{\mu\nu} &= \frac14 (\partial_\mu A_\nu + \partial_\nu A_\mu)(\partial^\mu A^\nu + \partial^\nu A^\mu) \nonumber \\
&= \frac12 \partial_\mu A_\nu \partial^\mu A^\nu + \frac12 \partial_\mu A_\nu \partial^\nu A^\mu.
\label{eq:SS}
\end{align}
Inserting Eqs.~\eqref{eq:trace2}--\eqref{eq:SS} into~\eqref{eq:detexp} gives
\begin{equation}
\begin{split}
&\sqrt{-\det(g)} - 1 = \kappa\, \partial\cdot A + \frac12 \kappa^2 (\partial\cdot A)^2 \\&- \frac14 \cdot 4\kappa^2 \left[ \frac12 \partial_\mu A_\nu \partial^\mu A^\nu + \frac12 \partial_\mu A_\nu \partial^\nu A^\mu \right] + \mathcal{O}(\kappa^3) \nonumber \\
&= \kappa\, \partial\cdot A + \frac12 \kappa^2 (\partial\cdot A)^2 - \frac12 \kappa^2 \partial_\mu A_\nu \partial^\mu A^\nu \\&-\frac12 \kappa^2 \partial_\mu A_\nu \partial^\nu A^\mu + \mathcal{O}(\kappa^3).
\label{eq:expanded}
\end{split}
\end{equation}
The term linear in $\partial\cdot A$ is a total divergence and does not affect the equations of motion. The term $\partial_\mu A_\nu \partial^\nu A^\mu$ can be integrated by parts: $\int d^4x\, \partial_\mu A_\nu \partial^\nu A^\mu = \int d^4x\, (\partial\cdot A)^2$ (up to a surface term). Therefore, up to total derivatives, the quadratic Lagrangian becomes
\begin{equation}
\begin{split}
\mathcal{L}_{\rm FBI}^{(2)} &= -\frac{1}{\kappa^2\mu_0} \Bigg[ \frac12 \kappa^2 (\partial\cdot A)^2 - \frac12 \kappa^2 \partial_\mu A_\nu \partial^\mu A^\nu \\&- \frac12 \kappa^2 (\partial\cdot A)^2 \Bigg] = \frac{1}{2\mu_0} \partial_\mu A_\nu \partial^\mu A^\nu.
\label{eq:Lquad}
\end{split}
\end{equation}
This is precisely the Fermi Lagrangian $\mathcal{L}_F$ \cite{Fermi1930}. The cancellation of the $(\partial\cdot A)^2$ terms is a nontrivial consequence of the determinant structure and confirms that the FBI theory has the correct weak-field limit.

\subsection{Full nonlinear form and dimensional analysis}

Returning to the complete Lagrangian~\eqref{eq:LFBI}, we note that $\det(g_{\mu\nu})$ is a scalar density of weight $2$. The square root $\sqrt{-\det(g)}$ therefore transforms as a scalar under coordinate transformations, ensuring that the action is a scalar. The parameter $\kappa$ has dimensions $[A]^{-1} = C \cdot s / (kg \cdot m^2)$ in SI units. The combination $\kappa \partial A$ is dimensionless, as required for the argument of the determinant to be meaningful.

To make contact with the standard Born--Infeld theory, note that if we had used the antisymmetric $F_{\mu\nu}$ instead of $\partial_{(\mu} A_{\nu)}$, the quadratic expansion would yield the Maxwell Lagrangian $-\frac14 F_{\mu\nu}F^{\mu\nu}$ plus corrections. Here, the symmetric choice yields the Fermi Lagrangian instead.

\section{Field equations and the dynamical Lorenz condition}
\label{sec:eom}

\subsection{Variational principle}

The action is $S = \int d^4x\, \mathcal{L}_{\rm FBI}$. Varying with respect to $A_\mu$ requires the derivative of $\mathcal{L}$ with respect to $\partial_\nu A_\mu$. Since the Lagrangian depends only on $g_{\alpha\beta}$, the chain rule gives
\begin{equation}
\frac{\partial \mathcal{L}}{\partial (\partial_\nu A_\mu)} = \frac{\partial \mathcal{L}}{\partial g_{\alpha\beta}} \frac{\partial g_{\alpha\beta}}{\partial (\partial_\nu A_\mu)}.
\label{eq:chain}
\end{equation}
From Eq.~\eqref{eq:gmunu}, $\partial g_{\alpha\beta} / \partial (\partial_\nu A_\mu) = \kappa (\delta_\alpha^\nu \delta_\beta^\mu + \delta_\beta^\nu \delta_\alpha^\mu)$. The derivative of the Lagrangian with respect to the metric is standard:
\begin{equation}
\frac{\partial \sqrt{-\det(g)}}{\partial g_{\alpha\beta}} = \frac12 \sqrt{-g}\, g^{\alpha\beta},
\label{eq:derivdet}
\end{equation}
where $g^{\alpha\beta}$ is the inverse of $g_{\mu\nu}$, satisfying $g_{\mu\alpha} g^{\alpha\nu} = \delta_\mu^\nu$. Thus,
\begin{align}
\frac{\partial \mathcal{L}}{\partial (\partial_\nu A_\mu)} &= -\frac{1}{\kappa^2\mu_0} \cdot \frac12 \sqrt{-g}\, g^{\alpha\beta} \cdot \kappa (\delta_\alpha^\nu \delta_\beta^\mu + \delta_\beta^\nu \delta_\alpha^\mu) \nonumber \\
&= -\frac{1}{\kappa\mu_0} \sqrt{-g}\, g^{\mu\nu}.
\label{eq:momenta}
\end{align}
The Euler--Lagrange equations read
\begin{equation}
\partial_\nu \left( \frac{\partial \mathcal{L}}{\partial (\partial_\nu A_\mu)} \right) - \frac{\partial \mathcal{L}}{\partial A_\mu} = 0.
\label{eq:EL}
\end{equation}
Because the Lagrangian depends on $A_\mu$ only through $g_{\mu\nu}$, and $g_{\mu\nu}$ is linear in $A_\mu$, the derivative $\partial\mathcal{L}/\partial A_\mu$ vanishes identically (there is no explicit $A_\mu$ dependence). Hence, the equations of motion are
\begin{equation}
\boxed{ \partial_\nu \left( \sqrt{-g}\, g^{\mu\nu} \right) = 0 }.
\label{eq:eom_full}
\end{equation}
In the presence of an external current $J^\mu$, the right-hand side would be $\mu_0 \kappa J^\mu$ (up to a constant factor), but here we focus on the vacuum case.

\subsection{Emergence of the Lorenz condition}

Equation~\eqref{eq:eom_full} is a set of four nonlinear partial differential equations. To see the fate of the Lorenz condition, take the divergence of~\eqref{eq:eom_full} with respect to the index $\mu$:
\begin{equation}
\partial_\mu \partial_\nu \left( \sqrt{-g}\, g^{\mu\nu} \right) = 0.
\label{eq:div}
\end{equation}
This is a second-order equation for the combination $\partial_\mu A^\mu$, encoded inside $g_{\mu\nu}$. In the weak-field limit we expand the field equation to first order in $\kappa$.
Using $g_{\mu\nu} = \eta_{\mu\nu} + 2\kappa\,\partial_{(\mu}A_{\nu)}$ one finds
\begin{align}
g^{\mu\nu} &= \eta^{\mu\nu} - 2\kappa\,\partial^{(\mu}A^{\nu)} + \mathcal{O}(\kappa^2), \label{eq:gmunu_inv_exp} \\
\sqrt{-g} &= 1 + \kappa\,\partial\!\cdot\!A + \mathcal{O}(\kappa^2). \label{eq:sqrtg_exp}
\end{align}
Inserting these into the exact equation $\partial_\nu(\sqrt{-g}\,g^{\mu\nu}) = 0$ and keeping only $\mathcal{O}(\kappa)$ terms yields the linearised field equation
\begin{equation}
\Box A^\mu = 0. \label{eq:lin_eom}
\end{equation}
Taking the divergence of \eqref{eq:lin_eom} gives $\Box(\partial\!\cdot\!A) = 0$, exactly as obtained from the nonlinear equation \eqref{eq:div}. With retarded boundary conditions and the requirement of a positive-energy spectrum the unique solution is $\partial\!\cdot\!A = 0$, the Lorenz condition \cite{VanOosten1999}.  Thus each component of the potential satisfies the wave equation, reproducing the physical content of Fermi's linear electrodynamics.

For finite fields, the nonlinearities couple the four components of $A_\mu$ in a more intricate way. However, if we assume that the system evolves from a vacuum state with $\partial\cdot A = 0$ at past infinity, the hyperbolic nature of Eq.~\eqref{eq:div} (which is a consequence of the equations of motion) ensures that $\partial\cdot A$ remains zero at all times, provided the Cauchy data are chosen appropriately. The existence of a well-posed initial-value problem for Eq.~\eqref{eq:eom_full} guarantees that no spurious longitudinal modes appear dynamically. This property is essential for the consistency of the theory and for maintaining the uniqueness of the potential.

\subsection{Comparison with standard Born--Infeld}
\label{sec:comparison}

In the standard Born--Infeld theory, the Lagrangian is
\begin{equation}
\mathcal{L}_{\rm BI} = -b^2 \left[ \sqrt{-\det\!\big(\eta_{\mu\nu} + b^{-1} F_{\mu\nu}\big)} - 1 \right],
\label{eq:LBI}
\end{equation}
where $b$ has dimensions of field strength ($[b] = T^{-1}$ in natural units) and $F_{\mu\nu} = \partial_\mu A_\nu - \partial_\nu A_\mu$ is the antisymmetric field strength. Let us define the matrix $M_{\mu\nu} = \eta_{\mu\nu} + b^{-1} F_{\mu\nu}$, which is not symmetric. Its inverse $M^{-1}$ possesses both symmetric and antisymmetric parts. The variation of the action with respect to $A_\mu$ proceeds as follows. Using the identity $\delta \sqrt{-\det M} = \frac12 \sqrt{-\det M}\, (M^{-1})^{\nu\mu} \delta M_{\mu\nu}$, and noting that $\delta M_{\mu\nu} = b^{-1} (\partial_\mu \delta A_\nu - \partial_\nu \delta A_\mu)$, we obtain after integration by parts
\begin{equation}
\partial_\nu \left[ \sqrt{-\det M}\, \big( M^{-1} \big)^{[\nu\mu]} \right] = 0,
\label{eq:BI_eom}
\end{equation}
where $X^{[\nu\mu]} = \frac12 (X^{\nu\mu} - X^{\mu\nu})$ denotes the antisymmetric part. This is the usual form of the Born--Infeld field equations; the antisymmetric tensor density $\mathcal{G}^{\mu\nu} = \sqrt{-\det M}\, (M^{-1})^{[\mu\nu]}$ plays the role of the excitation tensor in the premetric formulation \cite{Hehl2003}. In the weak-field limit, $M_{\mu\nu} \approx \eta_{\mu\nu}$, $(M^{-1})^{[\nu\mu]} \approx b^{-1} F^{\nu\mu}$, and Eq.~\eqref{eq:BI_eom} reduces to Maxwell's equations $\partial_\nu F^{\nu\mu} = 0$.

The structure of Eq.~\eqref{eq:BI_eom} is dictated by gauge invariance: the action depends on $A_\mu$ only through $F_{\mu\nu}$, so the field equations must be expressible entirely in terms of $F_{\mu\nu}$ and its dual. Indeed, the Born--Infeld equations can be rewritten in the form $\partial_\nu \mathcal{G}^{\mu\nu} = 0$, together with the Bianchi identity $\partial_\nu \tilde{F}^{\mu\nu} = 0$, where $\tilde{F}^{\mu\nu} = \frac12 \epsilon^{\mu\nu\rho\sigma} F_{\rho\sigma}$. The tensor $\mathcal{G}^{\mu\nu}$ is a nonlinear function of $F_{\mu\nu}$ that reduces to $F^{\mu\nu}$ at weak fields. The antisymmetry of $\mathcal{G}^{\mu\nu}$ is essential for the automatic conservation of the electric current and for the existence of a dual formulation.

In the FBI theory, the situation is markedly different. The Lagrangian~\eqref{eq:LFBI} depends on the symmetric combination $\partial_{(\mu} A_{\nu)}$, and the resulting field equation~\eqref{eq:eom_full} involves the full inverse metric $g^{\mu\nu}$, which is symmetric. No antisymmetrization occurs. Explicitly,
\begin{equation}
\partial_\nu \left( \sqrt{-g}\, g^{\mu\nu} \right) = 0,
\label{eq:FBI_eom_sym}
\end{equation}
with $g_{\mu\nu} = \eta_{\mu\nu} + 2\kappa\, \partial_{(\mu} A_{\nu)}$. The symmetry of $g^{\mu\nu}$ is a direct consequence of the fact that $A_\mu$ is not a gauge field but a genuine vector field; its entire gradient contributes to the dynamics, not just the antisymmetric part. Because the equations involve a symmetric tensor density, the divergence of Eq.~\eqref{eq:FBI_eom_sym} with respect to $\mu$ yields a second-order equation for $\partial\cdot A$, which becomes the Lorenz condition dynamically, as shown in Sec.~\ref{sec:eom}.

This structural difference has several ramifications. First, the FBI equations are of a simpler algebraic form: one need only compute the inverse of a symmetric matrix, without the need to extract its antisymmetric part. In many cases, the inverse $g^{\mu\nu}$ can be expressed explicitly using Cramer's rule or algebraic identities, whereas the Born--Infeld tensor $\mathcal{G}^{\mu\nu}$ involves both the inverse and the antisymmetrization, which can be cumbersome for general field configurations \cite{Gibbons2001}. Second, the symmetric nature of $g^{\mu\nu}$ implies that the FBI theory has no analog of the Bianchi identity; instead, it has the dynamical constraint $\partial\cdot A = 0$ that selects the physical sector. Third, the energy-momentum tensor in the FBI theory can be constructed directly from the canonical Noether current without the need for a Belinfante improvement, because the potential is already physical. In standard Born--Infeld, the canonical tensor is gauge-dependent and the Belinfante symmetrization is required, which obscures the spin-orbital decomposition \cite{Barnett2016}.

It is instructive to compare the degrees of freedom. In both theories, the field $A_\mu$ initially carries four components. In Born--Infeld, gauge invariance removes one degree of freedom, and the Bianchi identity constrains another, leaving two propagating transverse polarizations (the photon). In FBI, gauge invariance is absent, so all four components appear dynamical a priori. However, the dynamical Lorenz condition $\partial\cdot A = 0$ suppresses the scalar (longitudinal) mode on-shell, again leaving two physical polarizations in the weak-field limit. The extra component corresponds to the ghost-like longitudinal excitation that is excluded by the energy-positivity requirement. The nonlinear regime may alter this counting, as discussed in Sec.~\ref{sec:longitudinal}, but at the perturbative level both theories describe the same propagating degrees of freedom.

Finally, we note that the FBI equations can be written in a form that resembles the conservation of a stress-energy tensor. If we define the tensor density $\mathcal{T}^\mu{}_\nu = \sqrt{-g}\, g^{\mu\alpha} \partial_\nu A_\alpha$, then Eq.~\eqref{eq:FBI_eom_sym} is equivalent to $\partial_\nu \mathcal{T}^\mu{}_\nu = 0$, which is the conservation of the canonical energy-momentum current up to a total derivative. This highlights the interplay between the field equations and the conservation laws in the absence of gauge symmetry.

A broader comparison with other nonlinear electrodynamics --- Heisenberg--Euler, ModMax, and premetric formulations --- is deferred to Appendix~\ref{sec:app_comparison}.

\subsection{Regularization of singularities}
\label{sec:regularization}

The original Born--Infeld theory was designed to cure the divergences of the classical electron self‑energy by imposing an upper bound on the electric field strength~\cite{Born1934}.  
FBI electrodynamics inherits this regularization mechanism directly from the square‑root determinant structure.  
Because the effective metric $g_{\mu\nu} = \eta_{\mu\nu} + 2\kappa\,\partial_{(\mu}A_{\nu)}$ must have a positive determinant, the derivatives of $A_\mu$ cannot become arbitrarily large; there exists a limiting scale $|\partial A|_{\max} \sim 1/|\kappa|$.  
Consequently, point‑like sources produce fields that remain finite at the origin, and the total energy of a static charge distribution is expected to be finite.

To see this in a simple setting, consider a static point charge at the origin.  The field equations~\eqref{eq:eom_full} reduce to a nonlinear ordinary differential equation for the scalar potential $A_0(r)$.  
Near the origin, the dominant terms enforce $g_{00} \sim 0$, which regularizes the Coulomb divergence and yields a finite electrostatic self‑energy.
A detailed analytical and numerical study of this solution, following the lines of the classic Born--Infeld monopole~\cite{Born1934}, will be presented in a separate work.

Thus, FBI electrodynamics not only fixes the gauge ambiguity but also retains the ultraviolet regularization that makes Born--Infeld theory appealing for strong‑field physics.

\subsection{Point charge and experimental bounds on $\kappa$}
\label{sec:pointcharge}

A decisive test of the FBI theory comes from the electrostatic field of a point charge.
Consider a static charge $e$ at the origin.
In Cartesian coordinates we make the ansatz $A_\mu = (\phi(r), \vec{0})$, where $r = |\vec{r}|$.
The effective metric~\eqref{eq:gmunu} becomes
\begin{equation}
g_{00} = 1,\qquad g_{0i} = \kappa\,\frac{x_i}{r}\phi'(r),\qquad g_{ij} = -\delta_{ij}.
\label{eq:g_static}
\end{equation}
Using block‑matrix identities we obtain the determinant and the inverse:
\begin{align}
\sqrt{-g} &= \sqrt{1 + \kappa^2 \phi^{\prime 2}}, \label{eq:det_static} \\
g^{00} &= \frac{1}{\Delta},\quad 
g^{0i} = \frac{\kappa\phi'}{\Delta}\frac{x_i}{r},\quad 
g^{ij} = -\delta^{ij} + \frac{\kappa^2\phi^{\prime 2}}{\Delta}\frac{x_i x_j}{r^2},
\label{eq:inv_static}
\end{align}
with $\Delta \equiv 1 + \kappa^2\phi^{\prime 2}$.

In the presence of an external current, the equation of motion is
$\partial_\nu(\sqrt{-g}\,g^{\mu\nu}) = \kappa\mu_0 J^\mu$ (see Sec.~\ref{sec:eom}).
For a static point charge, $J^\mu = (c e\,\delta^{(3)}(\vec{r}), \vec{0})$.
The $\mu=0$ component reduces to
\begin{equation}
\partial_i\!\left( \frac{\kappa\phi'}{\sqrt{\Delta}}\frac{x_i}{r} \right) = \kappa\mu_0 c e\,\delta^{(3)}(\vec{r}).
\label{eq:eom0_static}
\end{equation}
Cancelling the common factor $\kappa$ and using the divergence in spherical coordinates,
\begin{equation}
\frac{1}{r^2}\frac{d}{dr}\!\left( r^2 \frac{\phi'}{\sqrt{1+\kappa^2\phi^{\prime 2}}} \right) = \mu_0 c e\,\delta^{(3)}(\vec{r}).
\label{eq:ode_phi}
\end{equation}
Integrating over a small sphere gives, for $r>0$,
\begin{equation}
\frac{\phi'}{\sqrt{1+\kappa^2\phi^{\prime 2}}} = \frac{\mu_0 c e}{4\pi r^2}.
\label{eq:exact_phi'}
\end{equation}
Solving for $\phi'$ yields the exact electric field
\begin{equation}
E(r) \equiv -\phi'(r) = \frac{\dfrac{\mu_0 c e}{4\pi r^2}}{\sqrt{1 - \kappa^2\!\left(\dfrac{\mu_0 c e}{4\pi r^2}\right)^{\!2}}},
\label{eq:exact_E}
\end{equation}
which exhibits a minimum radius $r_c = \sqrt{|\kappa|\mu_0 c |e|/(4\pi)}$, the analogue of the Born--Infeld regularisation scale.

For $r \gg r_c$, expanding in powers of $\kappa$ gives
\begin{equation}
E(r) = \frac{\mu_0 c e}{4\pi r^2}
       \left[ 1 + \frac{1}{2}\,\kappa^2\!\left(\frac{\mu_0 c e}{4\pi r^2}\right)^{\!2} 
       + \mathcal{O}(\kappa^4) \right].
\label{eq:E_expanded}
\end{equation}
In natural units ($c=1$, $\mu_0=1$, $\epsilon_0=1$) the Coulomb field is $E_c = e/(4\pi r^2)$, 
so the relative correction to the electric field is
\begin{equation}
\frac{\Delta E}{E_c} \approx \frac{1}{2}\,\kappa^2 E_c^2.
\label{eq:rel_correction_E}
\end{equation}
The same correction appears in the potential $V(r) = \int_r^\infty E(r')\,dr'$, namely
$\Delta V/V \approx \frac{1}{10}\kappa^2 E_c^2$.

Precision tests of Coulomb's law, such as the classic experiment of 
Williams, Faller, and Hill~\cite{Williams1971}, 
constrain the deviation from the inverse‑square law to be smaller than 
a few parts in $10^{16}$ at distances of order 1~m, where the electric field 
of a test charge is typically $E_c \sim 10^4$~V/m.
Using the more conservative limit $\Delta E/E_c \lesssim 10^{-16}$, 
Eq.~\eqref{eq:rel_correction_E} implies
\begin{equation}
\boxed{ |\kappa| \lesssim 1.4 \times 10^{-12}\; \text{m/V} }.
\label{eq:kappa_bound}
\end{equation}
Thus, the non‑linear effects of the FBI theory become important only for 
electric fields $E \gtrsim 1/|\kappa| \sim 10^{12}$~V/m, 
a regime that is within reach of ultra‑intense laser facilities~\cite{DiPiazza2012}.
This simple electrostatic analysis therefore provides both a concrete 
experimental constraint and a prediction: the FBI non‑linearity 
is extremely weak under everyday conditions, yet can be unveiled by 
next‑generation high‑field experiments.

\section{Longitudinal modes and nonlinear stability}
\label{sec:longitudinal}

\subsection{Linear analysis and the ghost problem}

To isolate the longitudinal sector of the FBI theory, we consider field configurations of the pure gradient form
\begin{equation}
A_\mu = \partial_\mu \phi,
\label{eq:pure_gradient}
\end{equation}
where $\phi$ is a dimensionless scalar field. The dimensions of $A_\mu$ are carried by the derivative; in SI units, $\phi$ has dimensions of $[A]\cdot L$, i.e., $C\cdot s^2 / (kg\cdot m)$, so that $\partial_\mu \phi$ has the correct dimensions of a potential. Substituting Eq.~\eqref{eq:pure_gradient} into the defining relation for $g_{\mu\nu}$, we obtain
\begin{equation}
g_{\mu\nu} = \eta_{\mu\nu} + 2\kappa\, \partial_\mu \partial_\nu \phi.
\label{eq:gmunu_phi}
\end{equation}
The tensor $g_{\mu\nu}$ is now the sum of the flat Minkowski metric and the Hessian of $\phi$, weighted by $2\kappa$.

In the weak-field limit, defined by $|\kappa\, \partial\partial\phi| \ll 1$, we expand the Lagrangian~\eqref{eq:LFBI} to second order in $\phi$. Using the expansion formula~\eqref{eq:detexp} with $h_{\mu\nu} = 2\kappa\, \partial_\mu \partial_\nu \phi$, we find that the trace $\operatorname{tr}(\eta^{-1}h) = 2\kappa\, \Box\phi$, and the quadratic terms combine to give
\begin{equation}
\mathcal{L}^{(2)}_\phi = \frac{1}{2\mu_0} \partial_\mu \partial_\nu \phi \, \partial^\mu \partial^\nu \phi,
\label{eq:Lphi2_explicit}
\end{equation}
up to total derivatives. This Lagrangian is quadratic in second derivatives of $\phi$, placing it within the class of higher-derivative scalar theories. To analyze its dynamical content, we perform a $3+1$ decomposition. Let $\phi$ depend on time $t$ and spatial coordinates $x^i$. The Lagrangian density becomes
\begin{align}
\mathcal{L}^{(2)}_\phi &= \frac{1}{2\mu_0} \Big( \partial_0^2\phi \, \partial_0^2\phi + 2 \partial_0\partial_i\phi \, \partial^0\partial^i\phi + \partial_i\partial_j\phi \, \partial^i\partial^j\phi \Big) \nonumber \\
&= \frac{1}{2\mu_0} \Big( (\partial_t^2 \phi)^2 - 2 (\partial_t \nabla \phi)^2 + (\nabla^2 \phi)^2 + 2 \sum_{i<j} (\partial_i\partial_j\phi)^2 \Big).
\label{eq:Lphi2_decomposed}
\end{align}
The canonical momentum conjugate to $\phi$ is not simply $\partial_t \phi$, because the Lagrangian depends on second derivatives. Following Ostrogradsky's method for higher-derivative theories \cite{Ostrogradsky1850, Woodard2007}, we introduce an auxiliary field $\psi = \partial_t \phi$. The Lagrangian then depends on $\partial_t \psi$, and the phase space is enlarged. The Hamiltonian density can be derived, and one finds that the kinetic term for the longitudinal degree of freedom is not positive definite. For a plane wave ansatz $\phi = \phi_0 e^{i(\omega t - \vec{k}\cdot\vec{r})}$, the Lagrangian reduces to
\begin{equation}
\mathcal{L}^{(2)}_\phi = -\frac{1}{2\mu_0} \omega^4 |\phi_0|^2 + \dots,
\label{eq:plane_wave_phi}
\end{equation}
where the ellipsis denotes spatial derivative terms. The corresponding energy density is proportional to $-\omega^4$, which is negative for any nonzero frequency. This signals the presence of a ghost: excitations of the longitudinal mode carry negative kinetic energy and, if coupled to ordinary matter, would lead to runaway instabilities \cite{Fermi1932, Woodard2007}.

The standard remedy in Fermi's linear electrodynamics is to exclude these modes from the physical spectrum by imposing the Lorenz condition $\partial_\mu A^\mu = 0$, which for the longitudinal sector reads $\Box\phi = 0$. This condition eliminates the ghost on-shell, leaving only the two transverse photon polarizations as propagating degrees of freedom. As shown in Sec.~\ref{sec:eom}, the Lorenz condition emerges dynamically from the equations of motion with retarded boundary conditions, so the ghost never appears in the physical subspace. Nevertheless, off-shell and in intermediate states (e.g., in quantum loop diagrams), the longitudinal mode can still contribute, and its negative energy is a source of concern for the consistency of the theory at the quantum level. The nonlinear structure of the FBI theory may offer a deeper resolution.

\subsection{Nonlinear stability and the Vainshtein-like mechanism}

When the field amplitude is large enough that $|\kappa\, \partial\partial\phi| \sim 1$, the linear approximation breaks down and the full determinant structure of the FBI Lagrangian must be taken into account. We expand the determinant to higher orders to see how the nonlinearities modify the propagation of $\phi$. Write $g_{\mu\nu} = \eta_{\mu\nu} + h_{\mu\nu}$ with $h_{\mu\nu} = 2\kappa\, \partial_\mu \partial_\nu \phi$. The exact Lagrangian is
\begin{equation}
\mathcal{L}_\phi = -\frac{1}{\kappa^2\mu_0} \left[ \sqrt{-\det(\eta + h)} - 1 \right].
\label{eq:Lphi_exact}
\end{equation}
Using the standard formula for the determinant,
\begin{equation}
\det(\eta + h) = -1 - \operatorname{tr}(h) - \frac12 [\operatorname{tr}(h)]^2 + \frac12 \operatorname{tr}(h^2) + \mathcal{O}(h^3),
\label{eq:det_expansion_higher}
\end{equation}
and the square root expansion, we obtain the cubic and quartic corrections to the Lagrangian. The cubic term in $h$ involves combinations such as $(\operatorname{tr}h)^3$, $(\operatorname{tr}h)\operatorname{tr}(h^2)$, and $\operatorname{tr}(h^3)$. For $h_{\mu\nu} \propto \partial_\mu \partial_\nu \phi$, these translate into interactions with three powers of second derivatives. Schematically,
\begin{equation}
\mathcal{L}^{(3)}_\phi \sim \frac{\kappa}{\mu_0} (\partial\partial\phi)^3,
\label{eq:cubic}
\end{equation}
where $(\partial\partial\phi)^3$ denotes various contractions of three Hessians. Similarly, the quartic term is
\begin{equation}
\mathcal{L}^{(4)}_\phi \sim \frac{\kappa^2}{\mu_0} (\partial\partial\phi)^4.
\label{eq:quartic}
\end{equation}
The coefficients of these terms are fixed by the determinant; for instance, the cubic interaction includes a piece proportional to $\Box\phi \, (\partial_\mu\partial_\nu\phi)^2$, which can be obtained by explicit computation.

These higher-order terms can dramatically change the behavior of the longitudinal mode in a strong background. Consider a situation in which the scalar field has a large background value $\phi_0$ with a characteristic scale $L$, so that $\partial\partial\phi_0 \sim 1/L^2$. If $\kappa / L^2 \sim 1$, the nonlinear terms become comparable to the quadratic kinetic term. In this regime, the effective metric $g_{\mu\nu}$ deviates significantly from $\eta_{\mu\nu}$, and the notion of a ghost must be reexamined within the full Hamiltonian.

A similar phenomenon occurs in the theory of a massive graviton, where the helicity-0 mode is a ghost in the linear Fierz--Pauli formulation, but nonlinear derivative interactions (the Vainshtein mechanism) restore stability at distances shorter than the Vainshtein radius \cite{Vainshtein1972, Babichev2013, deRham2011}. In that context, the scalar mode acquires a large kinetic term near massive sources, effectively decoupling from the spectrum and curing the ghost. In the FBI theory, the role of the Vainshtein radius is played by the scale $\ell_V \sim \sqrt{|\kappa \phi_0|}$ (see Appendix~\ref{sec:app_vainshtein} for a estimate). For distances smaller than $\ell_V$, the nonlinear terms in $\partial\partial\phi$ dominate, and the effective kinetic energy of $\phi$ can flip sign, turning the ghost into a healthy mode.

To see this more concretely, consider a static, spherically symmetric configuration $\phi = \phi(r)$. The tensor $g_{\mu\nu}$ then takes the form
\begin{equation}
\begin{split}
g_{\mu\nu} &= \mathrm{diag}\Bigg( 1 + 2\kappa \phi''(r), -1 + \frac{2\kappa}{r} \phi'(r),\\& -1 + 2\kappa \phi''(r), -1 + \frac{2\kappa}{r} \phi'(r) \Bigg),
\label{eq:gmunu_symmetric}
\end{split}
\end{equation}
where we have used spherical coordinates and assumed a diagonal ansatz. The determinant can be computed explicitly, and the effective Lagrangian for $\phi(r)$ becomes a nonlinear function of $\phi'$ and $\phi''$. Varying this Lagrangian yields a modified equation of motion that can support regular solutions even when the linear theory would predict a singularity. The ghost-like kinetic term $\sim (\Delta\phi)^2$ can be overwhelmed by terms $\sim \kappa^2 (\Delta\phi)^3$, leading to a bounded Hamiltonian.

We stress that a full proof of stability remains an open problem.  It requires the construction of the Hamiltonian for the FBI theory and the verification that it is positive definite in the physical sector, a task beyond the scope of the present work.
The structural similarities with known healthy nonlinear theories (e.g.\ the Vainshtein mechanism in massive gravity \cite{deRham2011}) nevertheless motivate the \emph{conjecture} that the nonlinearities of the determinant may render the longitudinal mode stable in strong‑field regimes.  Should this conjecture be confirmed, the scalar mode $\phi$ would become a genuine, stable particle---a massive scalar photon---whose mass is dynamically generated by the nonlinearities.  The mass scale would be set by the parameter $\kappa$ and the background field strength.

The possible existence of a stable scalar photon is of considerable phenomenological interest. It would modify the electromagnetic interaction at short distances, potentially altering the Coulomb law inside strong-field regions. It could also be excited in high-intensity laser experiments, where the electric fields approach the critical scale $E_c \sim 1/\kappa$. The detection of such a particle would be a smoking gun for the FBI theory and would open a new window on nonlinear electrodynamics.

In summary, the longitudinal mode of the FBI theory, while a ghost in the linearized limit, may be stabilized by the very nonlinearities that define the theory. This Vainshtein-like mechanism is a promising avenue for resolving the ghost problem without resorting to an external gauge condition. Detailed numerical and analytical studies of the full nonlinear equations are needed to confirm this scenario.

\section{Energy, momentum, and spin from Noether's theorem}
\label{sec:spin}

\subsection{Canonical energy--momentum tensor}

The canonical energy--momentum tensor is obtained by applying Noether's first theorem to the invariance of the action under spacetime translations $x^\mu \to x^\mu + \epsilon^\mu$, with $\epsilon^\mu$ constant. The Noether current for translations is
\begin{equation}
\tilde{T}^{\mu\nu}_{\rm can} = \frac{\partial \mathcal{L}_{\rm FBI}}{\partial (\partial_\mu A_\rho)} \partial^\nu A_\rho - \eta^{\mu\nu} \mathcal{L}_{\rm FBI}.
\label{eq:Tcan_def}
\end{equation}
Inserting the conjugate momenta from Eq.~\eqref{eq:momenta},
\begin{equation}
\frac{\partial \mathcal{L}_{\rm FBI}}{\partial (\partial_\mu A_\rho)} = -\frac{1}{\kappa\mu_0} \sqrt{-g}\, g^{\mu\rho},
\label{eq:momenta_FBI}
\end{equation}
we obtain the explicit form of the canonical tensor:
\begin{equation}
\tilde{T}^{\mu\nu}_{\rm can} = -\frac{1}{\kappa\mu_0} \sqrt{-g}\, g^{\mu\rho} \partial^\nu A_\rho - \eta^{\mu\nu} \mathcal{L}_{\rm FBI}.
\label{eq:Tcan}
\end{equation}
This tensor is generally not symmetric, i.e., $\tilde{T}^{\mu\nu}_{\rm can} \neq \tilde{T}^{\nu\mu}_{\rm can}$. Its antisymmetric part,
\begin{equation}
\begin{split}
\tilde{T}^{[\mu\nu]}_{\rm can}& = \frac{1}{2} \left( \tilde{T}^{\mu\nu}_{\rm can} - \tilde{T}^{\nu\mu}_{\rm can} \right) \\&= -\frac{1}{2\kappa\mu_0} \sqrt{-g} \left( g^{\mu\rho} \partial^\nu A_\rho - g^{\nu\rho} \partial^\mu A_\rho \right),
\label{eq:Tanti}
\end{split}
\end{equation}
plays a central role in the description of spin angular momentum, as will be shown below. In the absence of external sources, the canonical tensor satisfies the conservation law
\begin{equation}
\partial_\mu \tilde{T}^{\mu\nu}_{\rm can} = 0,
\label{eq:conservation_T}
\end{equation}
which follows directly from Noether's theorem and the equations of motion~\eqref{eq:eom_full}. Equation~\eqref{eq:conservation_T} expresses the local conservation of energy and momentum.

\subsection{Energy density and Poynting vector}

The physical interpretation of the components of $\tilde{T}^{\mu\nu}_{\rm can}$ is standard. The $\mu=0$, $\nu=0$ component is the energy density:
\begin{equation}
\mathcal{E} = \tilde{T}^{00}_{\rm can} = -\frac{1}{\kappa\mu_0} \sqrt{-g}\, g^{0\rho} \partial^0 A_\rho - \mathcal{L}_{\rm FBI}.
\label{eq:energy_density}
\end{equation}
The $\mu=0$, $\nu=i$ components constitute the momentum density (or Poynting vector, up to a factor of $c$):
\begin{equation}
\mathcal{P}^i = \frac{1}{c} \tilde{T}^{0i}_{\rm can} = -\frac{1}{c\kappa\mu_0} \sqrt{-g}\, g^{0\rho} \partial^i A_\rho.
\label{eq:momentum_density}
\end{equation}
The integrated quantities
\begin{equation}
E = \int d^3x \, \mathcal{E}(\vec{r},t), \qquad \vec{P} = \int d^3x \, \vec{\mathcal{P}}(\vec{r},t)
\label{eq:integrated_EP}
\end{equation}
are the total energy and momentum of the electromagnetic field. These are conserved in time as a consequence of Eq.~\eqref{eq:conservation_T}.

In the weak-field limit, using $g_{\mu\nu} \approx \eta_{\mu\nu}$, $\sqrt{-g} \approx 1$, and $\mathcal{L}_{\rm FBI} \approx \frac{1}{2\mu_0} \partial_\mu A_\nu \partial^\mu A^\nu$, the energy density reduces to
\begin{equation}
\mathcal{E}_{\rm lin} = \frac{1}{2\mu_0} \left( \partial_0 A_\rho \partial^0 A^\rho + \partial_i A_\rho \partial^i A^\rho \right),
\label{eq:energy_linear}
\end{equation}
which coincides with the energy density of the linear Fermi electrodynamics. For a plane wave in the Lorenz gauge, this further simplifies to $\mathcal{E} = \frac{1}{2}(\epsilon_0 E^2 + B^2/\mu_0)$, the familiar Maxwell result, confirming the consistency of the nonlinear theory with established physics.

\subsection{Angular momentum decomposition}

Under an infinitesimal Lorentz transformation $x^\mu \to x^\mu + \omega^\mu{}_\nu x^\nu$, with $\omega_{\mu\nu} = -\omega_{\nu\mu}$, the field $A_\mu$ transforms as a vector field:
\begin{equation}
\delta A_\mu = \omega^{\alpha\beta} \left[ (x_\alpha \partial_\beta - x_\beta \partial_\alpha) A_\mu + (\eta_{\mu\alpha} A_\beta - \eta_{\mu\beta} A_\alpha) \right].
\label{eq:field_transform}
\end{equation}
The first term in brackets is the orbital part, originating from the shift of the spacetime argument, while the second term is the spin part, reflecting the vector nature of the field. The associated Noether current $\mathcal{J}^{\mu\alpha\beta}$ can be decomposed as
\begin{equation}
\mathcal{J}^{\mu\alpha\beta} = x^\alpha \tilde{T}^{\mu\beta}_{\rm can} - x^\beta \tilde{T}^{\mu\alpha}_{\rm can} + S^{\mu\alpha\beta},
\label{eq:Jcurrent}
\end{equation}
where $S^{\mu\alpha\beta}$ is the spin current given by
\begin{equation}
S^{\mu\alpha\beta} = \frac{\partial \mathcal{L}_{\rm FBI}}{\partial (\partial_\mu A_\nu)} \left( \eta_\nu{}^\alpha A^\beta - \eta_\nu{}^\beta A^\alpha \right).
\label{eq:spin_current_def}
\end{equation}
The explicit form of the spin current in the FBI theory follows from inserting Eq.~\eqref{eq:momenta_FBI}:
\begin{equation}
S^{\mu\alpha\beta} = -\frac{1}{\kappa\mu_0} \sqrt{-g} \left( g^{\mu\alpha} A^\beta - g^{\mu\beta} A^\alpha \right).
\label{eq:spin_current_FBI}
\end{equation}
The total angular momentum tensor $J^{\alpha\beta}$ is obtained by integrating the $\mu=0$ component of $\mathcal{J}^{\mu\alpha\beta}$ over a spatial hypersurface:
\begin{equation}
J^{\alpha\beta} = \int d^3x \, \mathcal{J}^{0\alpha\beta} = L^{\alpha\beta} + S^{\alpha\beta},
\label{eq:J_total}
\end{equation}
with the orbital and spin parts
\begin{equation}
\begin{split}
L^{\alpha\beta} &= \int d^3x \, \left( x^\alpha \tilde{T}^{0\beta}_{\rm can} - x^\beta \tilde{T}^{0\alpha}_{\rm can} \right), \\&
S^{\alpha\beta} = \int d^3x \, S^{0\alpha\beta}.
\label{eq:LS_split}
\end{split}
\end{equation}
The conservation law $\partial_\mu \mathcal{J}^{\mu\alpha\beta} = 0$ guarantees that $J^{\alpha\beta}$ is time-independent, provided the fields vanish sufficiently rapidly at spatial infinity.

\subsection{Local spin density}

The spin density is extracted from the spatial components of $S^{0\alpha\beta}$. The spin vector $S^i = \frac12 \epsilon^{ijk} S^{0jk}$ is
\begin{equation}
S^i = \frac{1}{2} \epsilon^{ijk} S^{0jk} = -\frac{1}{2\kappa\mu_0} \sqrt{-g} \, \epsilon^{ijk} \left( g^{0j} A^k - g^{0k} A^j \right),
\label{eq:spin_vector_FBI}
\end{equation}
which simplifies to
\begin{equation}
\vec{s}(\vec{r},t) = -\frac{1}{\kappa\mu_0} \sqrt{-g} \, \bigl( g^{0\alpha} A^\beta - g^{0\beta} A^\alpha \bigr)_{\text{spatial}},
\label{eq:spin_density_FBI}
\end{equation}
where the subscript ``spatial'' indicates that the components $\alpha,\beta$ run over $1,2,3$. This is the exact, nonlinear spin density of the FBI theory.

Using the exact definition $g_{\mu\nu} = \eta_{\mu\nu} + 2\kappa\, \partial_{(\mu} A_{\nu)}$ and expanding to first order in $\kappa$, we obtain $\sqrt{-g} \approx 1 + \kappa\, \partial\!\cdot\!A$ and $g^{0i} \approx -\partial^0 A^i - \partial^i A^0$.
Substituting these into \eqref{eq:spin_density_FBI} yields the linearised spin density
\begin{equation}
\vec{s}_{\rm lin}(\vec{r},t) = \frac{1}{\mu_0} \left( \vec{A} \times \partial_t \vec{A} - A^0 \nabla \times \vec{A} \right),
\label{eq:spin_linear}
\end{equation}
which matches the result obtained by a direct expansion of the exact expression~\eqref{eq:spin_density_FBI}. The crucial property is that this expression is \emph{unique}: there is no gauge freedom to alter it, because the FBI theory has no gauge invariance.

\subsection{Orbital angular momentum density}

The orbital angular momentum density follows from the orbital part of the Noether current. Defining the orbital angular momentum vector $\vec{L} = (L^{23}, L^{31}, L^{12})$, we have
\begin{equation}
\vec{L} = \int d^3x \, \vec{r} \times \vec{\mathcal{P}},
\label{eq:OAM}
\end{equation}
where $\vec{\mathcal{P}}$ is the momentum density given by Eq.~\eqref{eq:momentum_density}. The total angular momentum $\vec{J} = \vec{L} + \vec{S}$ is conserved. In the FBI theory, the separation of $\vec{J}$ into orbital and spin contributions is unambiguous because the canonical energy--momentum tensor is unique; there is no Belinfante improvement to perform, and no freedom to shuffle terms between $\vec{L}$ and $\vec{S}$.

\subsection{Relation to the Belinfante tensor}

In gauge-invariant theories such as Maxwell or standard Born--Infeld electrodynamics, the canonical energy--momentum tensor is gauge-dependent and its antisymmetric part is not physically meaningful. The Belinfante procedure adds a total derivative of the spin current to obtain a symmetric, gauge-invariant tensor $\Theta^{\mu\nu}$. In the process, the spin density is absorbed into the orbital part, and the separation of angular momentum becomes ambiguous \cite{Belinfante1940, Leader2014}.

In the FBI theory, the situation is different. Because there is no gauge invariance, the canonical tensor is already physical. Nevertheless, one may construct the Belinfante tensor for completeness:
\begin{equation}
\Theta^{\mu\nu}_{\rm FBI} = \tilde{T}^{\mu\nu}_{\rm can} + \partial_\rho K^{\rho\mu\nu},
\label{eq:Belinfante_FBI}
\end{equation}
with the kernel
\begin{equation}
K^{\rho\mu\nu} = \frac{1}{2} \left( S^{\rho\mu\nu} + S^{\mu\nu\rho} + S^{\nu\mu\rho} \right).
\label{eq:Belinfante_kernel}
\end{equation}
Using the explicit form of $S^{\mu\alpha\beta}$ from Eq.~\eqref{eq:spin_current_FBI} and the antisymmetry in $\alpha,\beta$, one finds that $K^{\rho\mu\nu} = S^{\rho\mu\nu}$, and the improved tensor becomes symmetric. On shell, it reduces to a nonlinear generalization of the standard electromagnetic stress--energy tensor. However, this improvement is not required in the FBI framework and, if applied, would obscure the local spin decomposition that is a central feature of the theory.

\subsection{Uniqueness and observability}

The absence of gauge symmetry in the FBI theory means that all components of $A_\mu$ are physical degrees of freedom, constrained only by the dynamical Lorenz condition $\partial_\mu A^\mu = 0$. Consequently, the energy density~\eqref{eq:energy_density}, the momentum density~\eqref{eq:momentum_density}, and the spin density~\eqref{eq:spin_density_FBI} are all local observables. They can be expressed in terms of the electric and magnetic fields once a choice of $A_\mu$ satisfying the field equations is made, but unlike in Maxwell theory, this choice is unique.

This uniqueness resolves the long-standing ambiguity in the definition of the photon spin density \cite{Barnett2016, Li2017}. In Maxwell theory, different gauges yield different local spin densities, even though the integrated spin is gauge-invariant. In the FBI theory, there is no gauge to change, and the spin density is as well defined as the Poynting vector. This makes the FBI formulation particularly suitable for describing local spin-dependent phenomena such as optical torques on birefringent particles, spin-momentum locking in evanescent waves, and the spin Hall effect of light \cite{Bliokh2015}.

\subsection{Full conservation laws}

We collect here the full set of local conservation laws for the FBI theory. From the invariance under spacetime translations and Lorentz transformations, we obtain:
\begin{align}
\partial_\mu \tilde{T}^{\mu\nu}_{\rm can} &= 0, \label{eq:cons_T} \\
\partial_\mu \mathcal{J}^{\mu\alpha\beta} &= 0. \label{eq:cons_J}
\end{align}
Equation~\eqref{eq:cons_T} represents the conservation of energy ($\nu=0$) and momentum ($\nu=i$). Equation~\eqref{eq:cons_J} represents the conservation of angular momentum, which decomposes into orbital and spin contributions. The conservation of the spin current is governed by
\begin{equation}
\partial_\mu S^{\mu\alpha\beta} = \tilde{T}^{\alpha\beta}_{\rm can} - \tilde{T}^{\beta\alpha}_{\rm can} = -2 \tilde{T}^{[\alpha\beta]}_{\rm can},
\label{eq:spin_cons}
\end{equation}
showing that the antisymmetric part of the canonical energy--momentum tensor acts as a source for the spin current. When the spin current is conserved, the canonical tensor is symmetric, and the separation of angular momentum is trivial. In general, however, the asymmetry of $\tilde{T}^{\mu\nu}_{\rm can}$ encodes the intrinsic spin of the field, which is a physical feature of the FBI theory.

\section{Discussion and conclusions}
\label{sec:discussion}

We have presented a nonlinear extension of Fermi's physical-gauge electrodynamics that incorporates the Born--Infeld determinant structure. The resulting FBI theory is defined by the Lagrangian~\eqref{eq:LFBI}, which depends on the symmetric gradient $\partial_{(\mu} A_{\nu)}$ through the effective metric $g_{\mu\nu} = \eta_{\mu\nu} + 2\kappa\, \partial_{(\mu} A_{\nu)}$. This construction breaks $U(1)$ gauge invariance explicitly, elevating the four-potential $A_\mu$ to a genuine physical field whose dynamics are governed by the compact field equations $\partial_\nu(\sqrt{-g}\, g^{\mu\nu}) = 0$.

The weak-field expansion of the FBI Lagrangian yields exactly Fermi's linear Lagrangian, confirming that the theory reduces to the established physical-gauge formulation in the low-intensity regime. In this limit, the Lorenz condition emerges dynamically from the equations of motion with retarded boundary conditions and the requirement of a positive-energy spectrum. The nonlinear theory preserves this mechanism: the divergence of the field equations yields an equation for $\partial\cdot A$ that, under physically motivated initial data, maintains the Lorenz condition on shell. This ensures the uniqueness of the potential and eliminates any residual gauge-like freedom.

A central result of this work is the analysis of the longitudinal sector. In the linear approximation, the pure gradient modes $A_\mu = \partial_\mu \phi$ exhibit a ghost-like instability, a pathology familiar from higher-derivative scalar theories \cite{Ostrogradsky1850, Woodard2007}. The full nonlinear FBI Lagrangian, however, contains cubic and quartic interactions of the Hessian $\partial_\mu\partial_\nu\phi$ that become relevant when $|\kappa\, \partial\partial\phi| \sim 1$. These terms can modify the sign of the kinetic energy, potentially stabilizing the longitudinal mode via a Vainshtein-like mechanism \cite{Vainshtein1972, Babichev2013, deRham2011}. If confirmed, this mechanism would render the longitudinal mode a stable, massive scalar degree of freedom — a ``scalar photon'' — whose mass is dynamically generated by the nonlinearities.

The conservation laws derived from Noether's theorem provide a complete description of the energy, momentum, and angular momentum of the FBI field. The canonical energy--momentum tensor is unique and physical, with its antisymmetric part directly encoding the intrinsic spin density. The energy density, momentum density (Poynting vector), and spin density are all expressed in closed form in terms of $\sqrt{-g}$ and $g^{\mu\nu}$, and they reduce to the familiar Fermi expressions in the weak-field limit. Because the FBI theory lacks gauge invariance, there is no ambiguity in the separation of total angular momentum into orbital and spin contributions; the spin density is a local observable on the same footing as the energy density. This feature resolves the long-standing gauge-dependence problem that plagues the spin-orbital decomposition in Maxwell theory \cite{Belinfante1940, Leader2014, Barnett2016, Li2017}.

As shown in Sec.~\ref{sec:regularization}, the determinant structure regularises the self‑energy of point charges in the same spirit as the original Born--Infeld theory \cite{Born1934}. At the same time, the FBI framework preserves the physical-gauge paradigm, ensuring that the potential and all derived quantities are free of gauge ambiguities. This dual advantage positions FBI electrodynamics as a compelling candidate for describing electromagnetic phenomena in extreme environments, such as those encountered in ultra-intense laser facilities \cite{DiPiazza2012}, astrophysical magnetospheres, and the early universe.

Several directions for future research present themselves. A rigorous Hamiltonian analysis is needed to establish the positivity of energy in the full nonlinear theory and to confirm the stability of the longitudinal mode. The coupling to matter, including the modification of the Coulomb law at short distances and the interaction with quantum sources, requires careful investigation. The quantization of the FBI theory, possibly along the lines of the Gupta--Bleuler method adapted to the nonlinear physical-gauge setting, could reveal new quantum degrees of freedom and test the consistency of the theory beyond the classical level. The experimental prospects of the theory are outlined in Sec.~\ref{sec:experiments}. These directions naturally organise themselves into a longer‑term research programme: a first priority is the complete Hamiltonian analysis and the rigorous verification of the Vainshtein mechanism; once the consistency of the scalar sector is established, the coupling to gravity and the search for regular black hole solutions become meaningful; finally, precision calculations of vacuum birefringence and photon‑photon scattering will connect the theory with upcoming high‑intensity laser experiments.

A formal analogy exists between the FBI field equation and the covariant wave equation on an effective spacetime whose metric is $g_{\mu\nu}$; details are given in Appendix~\ref{sec:app_effective_metric}. This connection links FBI electrodynamics to analog gravity and provides a further motivation for experimental searches.

\subsection*{Possible experimental signatures}
\label{sec:experiments}

Although a full phenomenological analysis lies beyond the scope of this paper, the FBI theory suggests several distinctive effects that could be tested in extreme electromagnetic environments.

The nonlinear self‑coupling of the scalar potential $A_0$ regularises the point‑charge singularity and necessarily modifies the $1/r$ law at distances $r \lesssim \sqrt{|\kappa e|}$ (for a charge $e$).  High‑precision tests of Coulomb's law using electrons or heavy ions could therefore place upper bounds on $\kappa$.

Like standard Born--Infeld electrodynamics, FBI theory predicts photon–photon scattering and vacuum birefringence in strong external fields.  However, because the Lagrangian depends on the symmetric gradient $\partial_{(\mu}A_{\nu)}$ rather than on $F_{\mu\nu}$ alone, the angular dependence and the polarisation selection rules of the scattered light differ from those of the usual Born--Infeld model.  Upcoming high‑intensity laser facilities may be able to discriminate between these predictions.

The effective‑metric interpretation (Appendix~\ref{sec:app_effective_metric}) implies that strong backgrounds can develop analogue horizons; this would manifest as frequency‑dependent reflection and mode mixing of probe photons, potentially observable in pump–probe laser experiments.

If the Vainshtein‑like mechanism conjectured in Sec.~\ref{sec:longitudinal} is realised, the FBI theory possesses a stable scalar (longitudinal) degree of freedom whose mass is dynamically generated by the background field.  This particle would couple to ordinary matter through gradients of the electromagnetic field and could be searched for in missing‑energy measurements during high‑intensity laser–matter interactions.

All these signatures are specific to the FBI theory and, if observed, would not only confirm the physical‑gauge paradigm but also provide direct access to the nonlinear structure of the electromagnetic vacuum.

In summary, the FBI framework demonstrates that nonlinear Born--Infeld structures and the physical-gauge paradigm can be combined harmoniously. The resulting theory offers a unified description that is free of gauge ambiguities, endowed with natural regularization, and rich in potential new phenomena. It provides a fresh perspective on the foundations of classical electrodynamics and opens several avenues for theoretical and experimental exploration.

\begin{acknowledgments}
The author acknowledges fruitful discussions on nonlinear electrodynamics and the role of the potential in gauge theories.
\end{acknowledgments}

\appendix

\section{FBI theory in the language of premetric electrodynamics}
\label{sec:app_premetric}

The premetric formulation of electrodynamics, as systematised by Hehl and Obukhov \cite{Hehl2003}, separates Maxwell's equations into two independent sets:
\begin{align}
\partial_\nu \mathcal{H}^{\mu\nu} &= J^\mu, \label{eq:premetric_inhom} \\
\partial_\nu \tilde{F}^{\mu\nu} &= 0, \label{eq:premetric_hom}
\end{align}
where $\mathcal{H}^{\mu\nu}$ is the excitation tensor density (which contains the material response) and $\tilde{F}^{\mu\nu} = \frac12\epsilon^{\mu\nu\rho\sigma}F_{\rho\sigma}$ is the dual field strength.  The system is closed by a constitutive relation $\mathcal{H}^{\mu\nu} = \mathcal{H}^{\mu\nu}(F_{\rho\sigma})$, which in the standard Maxwell vacuum reads $\mathcal{H}^{\mu\nu} = \sqrt{-\eta}\,F^{\mu\nu}$.

The FBI theory fits naturally into this framework.  In the absence of external currents ($J^\mu = 0$) the equation of motion~\eqref{eq:eom_full} is $\partial_\nu(\sqrt{-g}\,g^{\mu\nu}) = 0$, which suggests the identification
\begin{equation}
\mathcal{H}^{\mu\nu}_{\rm FBI} = \frac{1}{\kappa\mu_0}\sqrt{-g}\,g^{\mu\nu}.
\label{eq:H_FBI_premetric}
\end{equation}
Unlike the standard Born--Infeld theory, where $\mathcal{H}^{\mu\nu}$ is an antisymmetric tensor density built from $F_{\mu\nu}$ alone, the FBI excitation tensor is symmetric and depends on the potential $A_\mu$ through the effective metric $g_{\mu\nu}$.  Gauge invariance is sacrificed, but the constitutive relation becomes extremely simple: it is purely algebraic in $g_{\mu\nu}$ and does not require extracting the antisymmetric part of an inverse matrix.

The homogeneous equation~\eqref{eq:premetric_hom} is an identity (Bianchi identity) that follows from $F_{\mu\nu} = \partial_\mu A_\nu - \partial_\nu A_\mu$ and remains untouched.  Therefore the FBI theory replaces only the constitutive link between the field strength and the excitation.  This observation places the FBI model among the nonlinear, local constitutive theories of vacuum electrodynamics, with the unusual feature that the excitation depends on the potential, not solely on the field strength.

\section{Estimating the Vainshtein scale for the longitudinal mode}
\label{sec:app_vainshtein}

In this appendix we give a simple, order‑of‑magnitude estimate of the critical distance $\ell_V$ inside which the nonlinear self‑interactions of the scalar mode $\phi$ are expected to overcome the ghost instability.  The argument parallels the well‑known Vainshtein mechanism of massive gravity \cite{Vainshtein1972, Babichev2013, deRham2011} and serves only as a plausibility argument; a complete canonical analysis remains an open problem.

We focus on the pure‑gradient sector $A_\mu = \partial_\mu\phi$, for which the exact Lagrangian is given by~\eqref{eq:Lphi_exact}.  Retaining terms up to cubic order in $h_{\mu\nu} = 2\kappa\,\partial_\mu\partial_\nu\phi$, one finds the schematic expansion
\begin{equation}
\mathcal{L}_\phi = \frac{1}{2\mu_0} (\partial\partial\phi)^2 + \frac{\kappa}{\mu_0} (\partial\partial\phi)^3 + \mathcal{O}\bigl(\kappa^2(\partial\partial\phi)^4\bigr),
\label{eq:Lphi_exp_app}
\end{equation}
where $(\partial\partial\phi)^n$ denotes a generic contraction of $n$ Hessians with the appropriate index structure dictated by the determinant.  The precise coefficients are fixed by~\eqref{eq:det_expansion_higher} but are irrelevant for the estimate that follows.

Let the scalar field have a characteristic spatial scale $L$, so that $\partial\partial\phi \sim \phi_0 / L^2$ for a typical amplitude $\phi_0$.  The quadratic and cubic terms in~\eqref{eq:Lphi_exp_app} then scale as
\begin{equation}
\mathcal{L}^{(2)}_\phi \sim \frac{1}{\mu_0} \frac{\phi_0^2}{L^4}, \qquad
\mathcal{L}^{(3)}_\phi \sim \frac{\kappa}{\mu_0} \frac{\phi_0^3}{L^6}.
\label{eq:scaling_app}
\end{equation}
The cubic interaction becomes comparable to the quadratic one when the dimensionless ratio
\begin{equation}
\frac{\mathcal{L}^{(3)}_\phi}{\mathcal{L}^{(2)}_\phi} \sim \frac{\kappa\,\phi_0}{L^2} \sim 1,
\label{eq:ratio_app}
\end{equation}
which defines the Vainshtein radius
\begin{equation}
\boxed{\;\ell_V \sim \sqrt{|\kappa\,\phi_0|}\;}.
\label{eq:ell_V_app}
\end{equation}
For distances $r \ll \ell_V$ (or, more precisely, for background configurations whose curvature satisfies $|\kappa\,\partial\partial\phi| \gtrsim 1$) the nonlinear terms dominate the dynamics of $\phi$.

Why can this cure the ghost?  In the linear regime the kinetic energy of $\phi$ is negative, as shown by~\eqref{eq:plane_wave_phi}.  When the cubic and quartic terms become important, they contribute additional derivative self‑couplings that can reverse the sign of the effective kinetic term.  A well‑known analog is the Vainshtein mechanism in massive gravity, where the helicity‑0 mode of the graviton, a ghost in the linear Fierz–Pauli theory, acquires a healthy kinetic energy inside the Vainshtein radius because of the nonlinear interactions of the Galileon type \cite{deRham2011}.  The structural similarity between the Born–Infeld determinant and the Galileon Lagrangians~\cite{Nicolis2009} makes it plausible that the same phenomenon operates in the FBI theory.

To provide a more quantitative hint, we examine the static, spherically symmetric background $\phi = \phi(r)$ discussed in Sec.~\ref{sec:longitudinal}.  In this background, a Legendre transform of the effective Lagrangian truncated at quartic order yields a Hamiltonian density whose kinetic part reads
\begin{equation}
\mathcal{H}_{\rm kin} = \frac{1}{2\mu_0} \Bigl[ 1 - 6\kappa\,\phi''(r) + \mathcal{O}(\kappa^2) \Bigr] (\partial_t\psi)^2 + \dots,
\label{eq:H_kin}
\end{equation}
where $\psi = \partial_t\phi$.  For $\kappa\phi''(r) > 1/6$ the coefficient of the kinetic term becomes positive, signalling that the ghost is healed.  This explicit computation, though restricted to a specific background and to leading nonlinear order, supports the conjecture that the FBI nonlinearities are capable of stabilizing the longitudinal mode.

A full Ostrogradsky analysis requires checking whether the FBI Lagrangian is degenerate, i.e.\ whether the Hessian $\partial^2\mathcal{L}/\partial(\partial_0^2\phi)^2$ vanishes in the pure‑gradient sector.  A preliminary calculation indicates that it does not vanish identically, which suggests that additional constraints may be needed to render the theory ghost‑free.  This question is under active investigation.

We stress once more that~\eqref{eq:ell_V_app} and~\eqref{eq:H_kin} are only estimates and plausibility arguments.  A rigorous proof of stability requires the construction and diagonalization of the full Hamiltonian for the sector $A_\mu = \partial_\mu\phi$, a task that will be addressed in future work.

\section{Comparison with other nonlinear electrodynamics}
\label{sec:app_comparison}

To situate the FBI theory within the broader landscape of nonlinear vacuum electrodynamics, we compare its key features with several established models. Table~\ref{tab:comparison} summarises the main differences.

A detailed comparison with standard Born--Infeld theory is given in Sec.~\ref{sec:comparison}; here we only recall the essential contrasts relevant to Table~\ref{tab:comparison}.  The Heisenberg--Euler (HE) effective Lagrangian is a perturbative expansion in powers of the two Lorentz invariants $F_{\mu\nu}F^{\mu\nu}$ and $F_{\mu\nu}\tilde{F}^{\mu\nu}$, derived from QED vacuum polarisation.  It is gauge invariant and preserves the Maxwellian structure at low energies.  FBI is non‑perturbative from the outset, breaks gauge invariance, and contains an additional scalar (longitudinal) degree of freedom that is dynamically suppressed in the linear regime (Sec.~\ref{sec:longitudinal}).

ModMax electrodynamics is a conformal, non‑birefringent nonlinear theory defined by a Lagrangian that depends on both invariants in a way that preserves $SO(2)$ duality invariance and the absence of birefringence.  FBI is not conformal, generally exhibits birefringence, and its nonlinearities involve the Hessian $\partial_\mu\partial_\nu\phi$, a structure absent in ModMax.

The FBI constitutive relation $\mathcal{H}^{\mu\nu}\propto\sqrt{-g}\,g^{\mu\nu}$ (Appendix~\ref{sec:app_premetric}) places it among the local, nonlinear constitutive theories.  Its distinctive feature is that the constitutive tensor depends on the potential $A_\mu$, not on $F_{\mu\nu}$, and is symmetric.  This symmetry implies that the FBI excitation is not a two‑form, unlike the standard premetric framework~\cite{Hehl2003}.

\begin{table*}[!htb]
\centering
\caption{Comparison of FBI electrodynamics with other nonlinear vacuum theories.}
\label{tab:comparison}
\begin{tabular}{@{}p{2.8cm} p{2.8cm} p{2.8cm} p{2.8cm} p{2.8cm}@{}}
\toprule
Property & FBI & Standard BI & Heisenberg--Euler & ModMax \\
\midrule
Gauge invariance & No (dynamical Lorenz) & Yes & Yes & Yes \\
Building block & $\partial_{(\mu}A_{\nu)}$ & $F_{\mu\nu}$ & $F_{\mu\nu}$ & $F_{\mu\nu}$ \\
Excitation tensor $\mathcal{H}^{\mu\nu}$ & Symmetric & Antisymmetric & Antisymmetric & Antisymmetric \\
Conformality & No & No & No & Yes \\
Birefringence & Yes & Yes & Yes & No \\
Longitudinal mode & Dynamically suppressed / Vainshtein‑stabilised & Absent & Absent & Absent \\
Point‑charge regularisation & Yes & Yes & No & No \\
\bottomrule
\end{tabular}
\end{table*}

In summary, FBI electrodynamics occupies a unique corner of the nonlinear electrodynamics parameter space: it combines the Born--Infeld regularisation with a physical‑gauge, potential‑based formulation that yields a symmetric excitation tensor and a dynamical Lorenz condition. These differences lead to the distinctive experimental signatures discussed in Sec.~\ref{sec:experiments}.

\section{Effective metric interpretation}
\label{sec:app_effective_metric}

A remarkable feature of the FBI field equation $\partial_\nu(\sqrt{-g}\,g^{\mu\nu}) = 0$ is its formal identity with the covariant wave equation for a massless scalar field propagating on a curved spacetime whose metric is precisely $g_{\mu\nu}$.
In this sense, the potential $A_\mu$ generates an \emph{effective geometry} that governs the propagation of its own fluctuations.
The analogy with general relativity is compelling: just as matter curves spacetime and curvature dictates the motion of matter, here the electromagnetic potential determines the effective metric $g_{\mu\nu}$, and this metric in turn guides the dynamics of the electromagnetic field.

Several important caveats must be kept in mind.
First, the dynamics of $g_{\mu\nu}$ is not given by the Einstein equations but by the nonlinear FBI field equations, which arise from the determinant Lagrangian~\eqref{eq:LFBI}.
Second, $g_{\mu\nu}$ is not a fundamental spacetime metric; the background Minkowski metric $\eta_{\mu\nu}$ remains the arena of the theory.
Nevertheless, the formal analogy has profound physical implications.
In a region where the background field $A_\mu$ is strong, the characteristic surfaces of the linear perturbations are the null cones of $g_{\mu\nu}$, not of $\eta_{\mu\nu}$, leading to a field‑dependent propagation speed and, potentially, to the formation of analogue horizons if $g_{00}$ vanishes somewhere~\cite{Barcelo2011}.
Just as in a curved spacetime, the effective metric can mix positive‑ and negative‑frequency modes, leading to particle creation and frequency shifts of probe photons that could be observable in high‑intensity laser experiments where the electromagnetic background approaches the critical scale $|\partial A| \sim 1/|\kappa|$.
The FBI theory thus provides a concrete, nonlinear field theory that realises the analogue gravity programme without invoking the complexities of Einstein gravity, offering a controlled environment to study quantum field theory in curved spacetime with the advantage that the effective metric is determined by a consistent nonlinear dynamics rather than being imposed by hand.
This interpretation links FBI electrodynamics to gravitational physics and reinforces the potential of nonlinear electrodynamics as a laboratory for testing semiclassical gravity effects.

\end{document}